\documentstyle[floats,prd,aps,epsf]{revtex} 

\newlength{\mathwidth}
\newlength{\leftover}
\newlength{\widthofstuff}
\def\mathflushleft#1{\setlength{\leftover}{\mathwidth}
\settowidth{\widthofstuff}{$\displaystyle #1$}
\addtolength{\leftover}{-\widthofstuff}
#1 \hskip\leftover}
\def\mathflushright#1{\setlength{\leftover}{\mathwidth}
\settowidth{\widthofstuff}{$\displaystyle #1$}
\addtolength{\leftover}{-\widthofstuff}
\hskip\leftover #1}
\def\mathcenter#1{\setlength{\leftover}{\mathwidth}
\settowidth{\widthofstuff}{$\displaystyle #1$}
\addtolength{\leftover}{-\widthofstuff}
\divide\leftover by 2
\hskip\leftover #1 \hskip\leftover}
\def\mathonefourth#1{\setlength{\leftover}{\mathwidth}
\settowidth{\widthofstuff}{$\displaystyle #1$}
\addtolength{\leftover}{-\widthofstuff}
\divide\leftover by 4
\hskip\leftover #1 \hskip\leftover\hskip\leftover\hskip\leftover}
\def\maththreefourths#1{\setlength{\leftover}{\mathwidth}
\settowidth{\widthofstuff}{$\displaystyle #1$}
\addtolength{\leftover}{-\widthofstuff}
\divide\leftover by 4
\hskip\leftover\hskip\leftover\hskip\leftover #1 \hskip\leftover}
\def\mathstrutter#1{\vrule width 0pt height #1}
\def\ms15{\mathstrutter{15pt}}
\def\mh{M}
\def\mc{m}
\def\rc{R}

\begin{document}
\twocolumn[\hsize\textwidth\columnwidth\hsize\csname@twocolumnfalse\endcsname

\title{Binary-Induced Collapse Of A Compact, Collisionless Cluster}

\author{
\begin{tabular}{c}
 \\
Matthew D. Duez, Eric T. Engelhard, John M. Fregeau, Kevin M. Huffenberger \\
and \\
Stuart L. Shapiro \\
  \\
\end{tabular}
}
\address{Departments of Physics and Astronomy, \& NCSA, University of Illinois at 
Urbana-Champaign, Urbana, IL 61801}

\maketitle

\begin{abstract}
We improve and extend Shapiro's~\cite{Us} model of a
relativistic, compact object which is stable in isolation but is
driven dynamically unstable by the tidal field of a binary companion.  
Our compact object consists of a dense swarm of test particles moving in 
randomly-oriented, initially circular, relativistic orbits about a nonrotating black hole. 
The binary companion is a distant, slowly inspiraling point mass. The tidal field of the 
companion is treated as a small perturbation on the background Schwarzschild geometry 
near the hole; the resulting metric is determined by solving the perturbation 
equations of Regge and Wheeler and Zerilli in the quasi-static limit. The perturbed 
spacetime supports Bekenstein's ~\cite{Bek} conjecture that the horizon area of a 
near-equilibrium black hole is an adiabatic invariant. We follow the evolution of the 
system and confirm that gravitational collapse can be induced in a compact collisionless 
cluster by the tidal field of a binary companion.

\end{abstract}

\vskip2pc]

\section{Introduction}

The possibility that massive neutron stars might be driven unstable to collapse
to black holes when placed in a close binary orbit was first suggested
by Wilson, Mathews and Maronetti (hereafter WMM; ~\cite{WWM}) on the basis
of their approximate relativistic numerical simulations.
This finding was quite unexpected, partly because it disagreed
with earlier Newtonian calculations ~\cite{LRS} which showed that tidal fields
stabilize binary stars against radial collapse. In fact, none of the follow-up
post-Newtonian (PN)~\cite{LAI2} ~\cite{LOM} or approximate
analytic analyses ~\cite{Kip} indicate
the presence of any relativistic radial instability in fluid binaries, 
nor does an independent dynamical simulation ~\cite{Shib}. 
No evidence is found for the WMM ``crushing effect'' in the relativistic
numerical calculations of either corotational ~\cite{Baum} or irrotational ~\cite{Bon}
binaries in quasi-equilibrium circular orbits. These later calculations are particularly
careful to adopt the same simplifications as WMM (e.g, a conformally flat 3-metric). 
All of the calculations except those of WMM suggest that the maximum allowed rest mass 
of a fluid star in a binary is in fact slightly larger than the value in isolation.

While it would appear that the WMM effect may not be present for fluid stars,
the work of WMM raises the interesting question as to whether
the ``crushing instability'' might exist for a different type of binary
system. Specifically, are there {\it any} relativistic binary systems for which tidal fields  
can trigger the collapse of a compact object known to be stable in isolation?   
To address this question, Shapiro ~\cite{Us} offered a simple candidate compact object
consisting of a test particle orbiting a Schwarzschild black
hole outside the innermost-stable circular orbit (ISCO). Such a ``compact object''
is obviously stable in isolation, but when it is placed in a binary 
orbit about a distant mass, an integration of the test-particle equations of motion reveals that
the tidal field of the distant mass can cause a test particle to plunge inside the hole. 

Shapiro's model is very promising as a simple illustration of the crushing instability
in action, but the original analysis, as he emphasized, was highly simplified and somewhat heuristic. 
First, the study was confined to two special test-particle orbits, one coplanar with and the other
perpendicular
to the companion orbital plane. Moreover, the motion of the particle in the perpendicular case
was artificially constrained to remain in a plane, precluding any precessional motion or wobbling. 
Most important, the analysis was based on a post-Newtonian treatment of the 3-body
problem in which the tidal piece of the equation for
the relative motion of the test particle about the black hole was treated to 
lowest (Newtonian) order and the nontidal piece replaced by the exact, fully relativistic  expression 
for geodesic motion in Schwarzschild geometry. While such a hybrid approach, similar in spirit to one
proposed by Kidder, Will and
Wiseman~\cite{LEK} for the 2-body problem, should capture the essential dynamics, it is
far from rigorous. In particular, it is not clear {\it a priori}
how reliable it is to use a Newtonian tidal term 
when the test particle is moving at high velocity in a very strong gravitational field close
to a black hole.

In this paper we improve and extend the simple model presented in ~\cite{Us}
of a compact object subject to binary-induced collapse. 
We use the equations of Regge and Wheeler ~\cite{RW} and Zerilli ~\cite{Z}
to derive the perturbation to the 
spacetime close to a Schwarzschild black hole due to a distant mass.  We note in
passing that
the resulting perturbed spacetime provides another example supporting the recent 
conjecture of Bekenstein ~\cite{Bek} that the
horizon area of a perturbed black hole is an adiabatic invariant. We obtain the geodesic
equations of motion in the perturbed spacetime and solve them to track the dynamical evolution 
of a dense spherical swarm of 20,000 test particles placed around the hole outside the
ISCO. The `black hole $+$ swarm' constitutes
a compact collisionless cluster whose stability against tidally-induced collapse we determine.
We compare our refined treatment with the simpler version presented in ~\cite{Us} and show that
the original equations track the behavior reasonably well.
We confirm that the crushing instability can occur in a binary system containing simple compact 
collisionless clusters like the ones we construct.

\section{HYBRID -- PN TREATMENT}
\subsection{Basic Equations}

Consider first a system of three bodies.  Assume that two of the 
bodies, 1 and 2, say, are much closer to each other than they are to the third, so that 
the influence of body 3 on the relative orbit of 1 and 2 may be treated as a small 
tidal perturbation.  
Specialize to the case in which body 1 is a Schwarzschild 
black hole of mass $\mh$ and body 2 is a test particle ($m_t \ll \mh $).  
Since it is far away from the 1-2 pair (the ``compact object''), body 3 can be 
treated as a point particle, with mass $\mc$.  
Let {\bf r} be the coordinate position of the test particle relative to the  
hole and {\bf \rc} be the coordinate position of the distant point mass relative to the 
hole. Following Shapiro ~\cite{Us}, we may write the equation for the relative motion 
of the test particle about the hole as 

\begin{equation}
\label{six}
\ddot{\bf r}  =\ -\ \left({\mh \over r^2}\right)\ [A{\bf e_r} + B{\bf v}] 
+ {\mc \over \rc^3} \ \left[ {3{\bf \rc} \cdot {\bf r} \over \rc^2}
\ \ {\bf \rc}\ \ -\ \ {\bf r} \right]\, .
\end{equation}
where ${\bf e_r} = {\partial\over\partial r}$.

In the Newtonian limit,
$A=1$ and $B=0$ in Eq.~(\ref{six}).  For 
isolated binaries, Lincoln and Will~\cite{CWL} derive
post-Newtonian expressions for $A$ and $B$ for arbitrary
masses, correct through 2.5PN order.  Kidder, Will and
Wiseman~\cite{LEK} provide a ``hybrid'' set
of equations in which the sum of the terms in $A$ and $B$
that are independent of the ratio
$\eta = \mu /(\mh +m_t), \mu = \mh m_t/(\mh +m_t) ,$ is replaced by the exact
expression for geodesic motion in the Schwarzschild geometry around
a body of mass $\mh$, while  the terms dependent on
$\eta$ are left unaffected. Their resulting equation  of
motion is therefore exact in the test-body limit $(\eta
\rightarrow 0)$  and is valid to 2.5PN order when
appropriately expanded for arbitrary masses. We shall
utilize these same hybrid expressions for $A$ and $B$ and,
for simplicity, work in the test-body limit by taking one
member of our close pair to have a mass much smaller than
the other. Adopting harmonic (de Donder) coordinates, the
resulting (Schwarzschild) expressions for $A$ and $B$ are
given by

\begin{equation}
\label{twelve}
A \ =\ \left[{1-\mh /r\over (1\ +\ \mh /r)^3}
\right]\ \ -\ \
\left[ {2-\mh /r \over 1-(\mh /r)^2}\right]\ {\mh \over r}\
\dot{r}^2\ +\ v^2 ,\end{equation}

\begin{equation}
\label{thirteen}
B\ =\ -\ \left[{4-2\mh /r \over
1-(\mh /r)^2}\right]\ \dot{r}\, ,
\end{equation} 
where

\begin{equation}
\label{fourteen}
v^2\ =\ \dot{r}^2\ +\ r^2 (\dot{\theta}^2 + \sin^2\theta\dot{\phi}^2).
\end{equation}

To model the three-body system, we must also know the 
position of the companion (body 3) relative to the compact object.  
The leading Newtonian piece of the equation of motion for {\bf R} 
gives

\begin{equation}
\label{five}
\ddot {\bf \rc}\ =\ -{\mh+\mc \over \rc^3}{\bf \rc} \, .
\end{equation}

We are interested in the dynamical behavior of the close
pair, regarded as a single ``compact object'', as it
inspirals toward the distant mass $\mc$ due to gravitational
radiation emission.  To treat the inspiral of this
``binary'' ($\mc$ in orbit about the ``compact object''),
we must include radiation reaction terms in the lowest order
(Newtonian) orbit Eq.~(\ref{five}).  Formally, such a
treatment requires a consistent expansion up to 2.5PN
order.   In lieu of this, we shall analyze the inspiral by
assuming that the binary is in a nearly circular, Keplerian
orbit, which undergoes a slow inspiral due to gravitational
radiation loss in the quadrupole limit. This assumption is
equivalent to inserting a quadrupole radiation reaction
potential in the binary orbit Eq.~(\ref{five}) and neglecting the
lower-order, (non-dissipative)  PN corrections in that
particular equation. While such an expression is not
formally consistent to 2.5PN order, it
faithfully tracks the  secular inspiral of the binary in the
limit treated here in which the binary system is at wide
(nonrelativistic) separation. The details of  the inspiral
are not important here, only that the inspiral serves to
bring a tidal perturber slowly in from infinity toward our
compact object~\cite{OUR}.  The resulting equations for the
binary inspiral are then ~\cite{Us}

\begin{equation}
\label{fifteen}
R(t)/R(0) = (1 -t/T)^{1/4}
\end{equation} and
\begin{equation} 
\label{sixteen}
\theta(t) -\theta(0) ={8 \over 5}\left({ \mh + \mc
\over R(0)^3}
\right)^{1/2}T [1 - (1-t/T)^{5/8}] \, ,
\end{equation} 
(Note the typo in ~\cite{Us}.)  The binary inspiral timescale $T$ is given by
\begin{equation}
\label{seventeen}
T/\mh = {5 \over 256} {(R(0)/\mh )^4
\over {((\mc+\mh)/\mh)(\mc/\mh)}} \, . 
\end{equation} 
We will use Eqs.~(\ref{fifteen})--(\ref{seventeen}) in analyzing the relative orbit  
Eq.~(\ref{six}).

\subsection{Numerical Implementation}
To assess the fate of the compact object, we followed the motion of
a spherical swarm of 20,000 test particles about the black hole.
At $t=0$, the test particles were placed randomly about the hole at a radius $r/\mh=5.9$,
well outside the ISCO of an isolated black hole at $r/\mh = 5$, and set in circular
orbits of arbitrary orientation.  We set $\mc = \mh$ and started 
the simulation with the companion at $\rc/\mh = 60$.  At this initial 
separation, the tidal field of the companion is negligible at the compact object, and the 
test particles begin in stable equilibrium orbits.  We integrate (\ref{six}) until 
the time at which $\rc/\mh = 26.30$.  Cartesian coordinates were used throughout.

The results of the simulation are summarized in Figure 1, where we plot the mean cluster 
radius as a function of companion separation.  
Clearly, the compact object 
remains stable until the companion comes within $\rc/\mh \approx 30$, at which point the
tidal field causes many of the particles to plunge into the hole.
We find that 
17,919 particles (89.6 \%) have fallen into the black hole by $t/\mh = 121,893$, the 
time at which the companion reaches $\rc/\mh = 26.30$. 
The simulation confirms the existence of a crushing instability.

\begin{figure}
\epsfxsize=3in
\begin{center}
\leavevmode
\epsffile{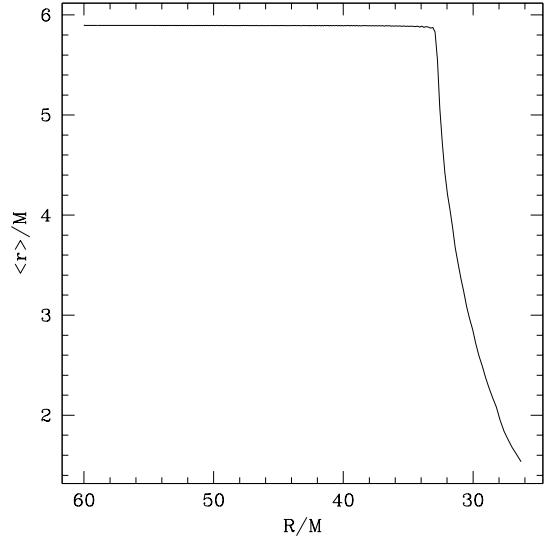}
\end{center}
\epsfxsize=3in
\begin{center}
\leavevmode
\epsffile{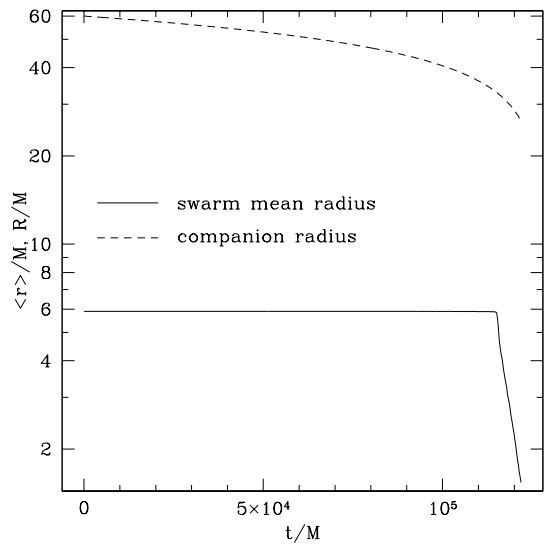}
\end{center}
\caption{Evolution of the cluster in the hybrid-PN approximation.
The mean radius of the 20,000 test-particles in the initially spherical swarm
is plotted as a function of the tidal companion radius in the top figure and
as a function of coordinate time in the bottom figure.}
\end{figure}

Several checks were performed to ensure the accuracy of our code, which 
integrates the ordinary differential orbit
equations by a standard fourth-order Runge Kutta algorithm with an adaptive stepsize ~\cite{NR}.
For example, in the absence of the companion, the code
conserves particle energy and angular momentum over the full time period of integration
and reliably locates the Schwarzschild ISCO with a controllable precision. These tests are
not so trivial when the integrations are performed in Cartesian coordinates.

\section{Schwarzschild Perturbation Treatment}
\subsection{Basic Equations}

\begin{figure}
\epsfxsize=1.5in
\begin{center}
\leavevmode
\epsffile{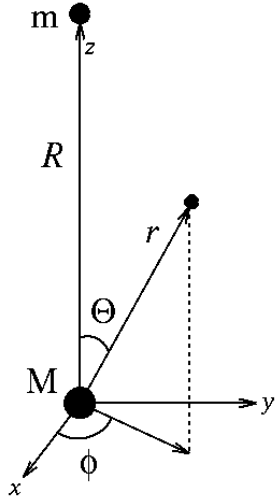}
\end{center}
\caption{The adopted coordinate system. The black hole $\mh$ is at the origin and the
companion $\mc$ is on the $z$ axis.}
\end{figure}

We will now improve the previous treatment by incorporating the tidal field of the 
companion star in a fully relativistic fashion.  To do this, we treat the effect of the companion as a 
small perturbation on a Schwarzschild background.  By assuming that the companion orbits at a much
greater distance from the black hole than the test particles, and hence moves with a much smaller angular
velocity, we can work in the quasi-static approximation.  We  
place the companion on the $z$ axis of our 
spherical polar coordinate system centered on the black hole (see Figure 2).
We follow Regge and Wheeler~\cite{RW} and divide the linear metric
perturbations into independent ``even'' and ``odd'' components
according to
\begin{equation}
\label{2one}
 g_{\mu\nu}\ =\ g_{\mu\nu}(S) + h_{\mu\nu}(\mbox{odd}) + h_{\mu\nu}(\mbox{even}) \, ,
\end{equation}
where $g_{\mu\nu}(S)$ is the usual Schwarzschild metric.
In the appropriate gauge, Regge and Wheeler found that the odd perturbations may be written as 
\begin{equation}
\label{2two}
\begin{array}{l}
\mathflushleft{h_{\mu\nu}(\mbox{odd})\ =\ \pmatrix{
0 & 0 & 0 & h_0(r) \cr
0 & 0 & 0 & h_1(r) \cr
0 & 0 & 0 & 0      \cr
h_0(r) & h_1(r) & 0 & 0 \cr}\times}\cr
\ms15 \mathflushright{\sin\theta {\partial \over \partial\theta}Y_{L0}(\theta)\times 
e^{(i\omega t)} \, .}\cr
\end{array}
\end{equation}
Likewise,  the even perturbations may be written, after simplifying gauge transformations, as
\begin{equation}
\label{2three}
\begin{array}{l}
\mathflushleft{h_{\mu\nu}(\mbox{even}) =}\\
\mathstrutter{35pt}\mathcenter{\pmatrix{\scriptstyle H_0(r)(1-2\mh/r) &\scriptstyle H_1(r) &\scriptstyle 0 &\scriptstyle 0 \cr
\scriptstyle H_1(r) & \mathstrutter{12pt}{\scriptstyle H_2(r) \over (1-2\mh/r)} & \scriptstyle 0 & \scriptstyle 0 \cr
\scriptstyle 0 & \scriptstyle 0 & \scriptstyle r^2 K(r) & \scriptstyle 0 \cr
\scriptstyle 0 & \scriptstyle 0 & \scriptstyle 0 & \scriptstyle r^2 \sin^2\theta K(r) \cr
}\times}\\
\mathstrutter{15pt}\mathflushright{Y_{L0}(\theta)\times e^{(i\omega t)} \, .}\\
\end{array}
\end{equation}
We now proceed 
to solve Einstein's equations to first order in the perturbation 
functions as in Zerilli ~\cite{Z}, using 
a delta function (point) source for the companion. We first find 
solutions to the homogeneous equations away from the companion 
and then use the inhomogeneous source terms to match these solutions.  
Setting $\omega = 0$ to comply with our quasi-static approximation, Zerilli's equations  
reduce to
\begin{equation}
\label{2four}
h_1 = H_1 = 0; \ \ \ \ \  H_0\ =\ H_2 \equiv H \, 
\end{equation}
\begin{equation}
\label{2five}
\begin{array}{l}
  \displaystyle
\mathflushleft{{d^2 h_0 \over dr^2} + \left[{4\mh \over r^2}-{L(L+1) \over r}\right]{h_0 \over r-2\mh} = }\\
\mathstrutter{15pt}\mathflushright{-8\pi r\left[{1 \over 2}L(L+1)\right]^{-1/2}Q_{L}^{(0)}\, ,}\\
\end{array}
\end{equation}
\begin{equation}
\label{2six}
\begin{array}{l}
\mathflushleft{\bigl(1 - {2\mh \over r}\bigr)^2 {d^2 K \over dr^2} + \bigl(1-{2\mh \over r}\bigr)\bigl(3-
{5\mh \over r}\bigr){1 \over r}{dK \over dr}}\\
\mathstrutter{15pt}\mathonefourth{- \bigl(1-{2\mh \over r}\bigr)^2\ {1 \over r}\ {dH \over dr}
  - \bigl(1-{2\mh \over r}\bigr){1 \over r^2}(H-K)}\\
\mathstrutter{15pt}\mathflushright{- \bigl(1-{2\mh \over r}\bigr) {1 \over 2r^2}L(L+1)(H+K) = 8\pi A_L^{(0)} \, ,}\\
\end{array}
\end{equation}
\begin{equation}
\label{2seven}
\mathcenter{-{1-\mh/r \over r-2\mh}{dK \over dr}+{1 \over r}{dH \over dr} + {2-L(L+1) \over 2r(r-2\mh)}(H-K) = 8\pi A_L \, .}
\end{equation}
In the above equations, $Q_L$, $A_L$, and $A_L^{(0)}$ represent components of the stress-energy tensor.
For a point source companion moving along a Schwarzschild geodesic, these terms are given in Appendix E of 
Zerilli ~\cite{Z}.  From 
Zerilli's tabulation of the source terms, one sees that all the source terms go to zero in the 
static limit except for $A_L^{(0)}$, i.e. $T_{00}$.

Homogeneous solutions may be found analytically for both the even and odd perturbations.  Regge and 
Wheeler show that, in the large $r$ limit (where the companion resides), there is one 
solution for $h_0$ varying like $r^{-L}$, and another varying like $r^{L+1}$.  Since there is no 
source term for $h_0$ ($Q_L = 0$), the only way that both $h_0$ and 
its first derivative can be continuous and regular everywhere is to require
$h_0 = 0$. Therefore, there is no contribution from odd parity 
perturbations.

The static solutions for $H$ and $K$ are found 
both by Regge and Wheeler~\cite{RW} and by Zerilli~\cite{Z} in terms of hypergeometric functions.  
We are interested in the lowest order nontrivial contribution, the quadrupole piece, 
generated by the companion, so 
we restrict our attention to the $L=2$ perturbation.  Inside the orbit of the companion, 
the solution which is regular at the horizon is 
\begin{equation}
\label{2eight}
\begin{array}{ll}
   \displaystyle
   H \ =\  \kappa_1 r(r-2\mh)  & \approx\kappa_1 r^2  \\
   \displaystyle
   K \ =\ \kappa_1 (r^2-2\mh^2) & \approx\kappa_1 r^2 \, ,
\end{array}
\end{equation}
where the last equality in each equation above holds at large $r \gg M$ and $\kappa_1$ is 
a constant to be determined by  
matching at the companion.  These equations may be verified by direct substitution into 
Eq.~(\ref{2six}) or (\ref{2seven}).  Outside the orbit of the companion, the perturbation is 
given by the solution regular at infinity,
\begin{equation}
\label{2nine}
\begin{array}{ll}
 H\ &=\ \kappa_2  \Bigl[{2\mh (2\mh^3+4\mh^2r-9\mh r^2+3r^3) \over (2\mh -r)r}  \\ 
    &  \hskip90pt +{3r^2(r-2\mh )^2\ln(1-2\mh /r) \over (2\mh -r)r}\Bigr] \\ 
    & \approx \kappa_2 r^{-3}  \\
    &     \\
 K\ &=\ \kappa_2  \mh^2\bigl[-6+{{4\mh}\over{r}}-{{6r}\over{\mh}}-{{3r^2\ln(1-2\mh /r)}\over{\mh^2}}  \\
    &   \ms15 \hskip 120pt + 6\ln(1-2\mh /r)\bigr] \\ 
    & \approx \kappa_2 r^{-3} \, ,
\end{array}
\end{equation}
where the last equality in each equation again holds at large $r \gg M$ and $\kappa_2$ is 
another constant to be determined by matching. 

Now we determine the two constants by matching the solutions at the radius of 
orbit of the companion, $\rc$.  
Because $A_L^{(0)}$, the source term of (\ref{2six}), does not go 
to zero in the quasi-static case, $\kappa_1$ and $\kappa_2$ do not vanish, as was the case for
the odd parity solutions.   
Taking $\rc \gg \mh$ as required by the quasi-static approximation, 
we use the asymptotic expressions in equations (\ref{2eight}) and (\ref{2nine}) for the matching. 
Requiring that $K$ be continuous across $\rc$ then yields our first condition
\begin{equation}
\label{2ten}
 \kappa_1 \rc^2 = \kappa_2 \rc^{-3} ~~\Rightarrow~~ \kappa_2 = \rc^5 \kappa_1 \, .
\end{equation}
The second condition may be found in either of two ways.  First, one may integrate Equation (\ref{2six}) 
across the source to obtain a
jump condition relating difference in the first derivative of $K$ on either side of $r=R$ to the 
source strength.  Alternatively, one may compute $\kappa_1$ directly by taking the Newtonian 
limit of $H$ and matching it to the Newtonian tidal potential.  Either way, one obtains
 $\kappa_1 = {{\mc}\over{\rc^3}} 4\sqrt{\pi/5}$.  Thus, $h_{00}$, for example, is 
\begin{equation}
\label{2thirteen}
\begin{array}{l}
\mathflushleft{h_{00} = (1-2\mh /r)(r^2-2\mh r)\kappa_1Y_{20}(\theta)}\\
\mathstrutter{15pt}\mathflushright{= (1-2\mh /r)(r^2 - 2\mh r){\mc \over \rc^3}(3\cos^2\theta - 1)\, .}\\
\end{array}
\end{equation}
From (\ref{2thirteen}), one can immediately verify that the metric derived above has the correct Newtonian limit---the 
potential reduces to the classic Hill potential.  This solution for the even parity solutions was first 
discovered by Moeckel~\cite{Mo}.

The entire metric may now be written out, defining $P \equiv P_2(\cos\theta) = 
3\cos^2\theta - 1$:
\begin{equation}
\label{2fourteen}
\begin{array}{l}
\mathflushleft{ds^2 = \Bigl[-\bigl(1-{2\mh \over r}\bigr)+{\mc \over \rc^3}P(r-2\mh )^2\Bigr]dt^2}\\
\mathstrutter{15pt}\mathcenter{+ \Bigl[\bigl(1-{2\mh \over r}\bigr)^{-1}+{\mc \over \rc^3}Pr^2\Bigr]dr^2}\\
\mathstrutter{15pt}\mathflushright{+ \Bigl[r^2 + {\mc \over \rc^3}Pr^2(r^2-2\mh^2)\Bigr]d\Omega^2 \, .}
\end{array}
\end{equation}
We note that the perturbed spacetime (\ref{2fourteen}) furnishes another example supporting the
conjecture of Bekenstein ~\cite{Bek} that the horizon area of a near-equilibrium
black hole is an adiabatic invariant (see Appendix).

To compare with the hybrid-PN treatment, we transform to harmonic coordinates 
by replacing the areal radial coordinate, $r$, with the harmonic radius, $r_h = r - \mh$.
Dropping the subscript, we obtain in harmonic coordinates
\begin{equation}
\label{2fifteen}
\begin{array}{l}
\mathflushleft{ds^2 \ = \displaystyle -{r-\mh \over r+\mh }dt^2 + {r+\mh \over r-\mh }dr^2 + (r+\mh)^2d\Omega^2}\\
\maththreefourths{\ms15 \displaystyle + {\mc \over \rc^3}P\bigl[(r-\mh)^2dt^2 + (r+\mh)^2dr^2} \\
\mathflushright{\displaystyle + (r+\mh )^2(r^2+2r\mh -\mh^2)d\Omega^2\bigr] \, .}\\
\end{array}
\end{equation}

Obtaining the geodesic equations in this spacetime 
up to first order in the tidal expansion parameter ${\mc r^2 \over \rc^3}$, 
we find the equations of motion for a test particle near the black hole to be
\begin{equation}
\label{2nineteen}
\begin{array}{l} 
  \displaystyle
\mathflushleft{\ddot r = {3\mh \dot r^2 \over r^2-\mh^2}-{\mh (r-\mh ) \over (r+\mh )^3} + (r-\mh )\Omega^2}\\
\mathonefourth{\ms15 + {\mc \over \rc^3} P \Big[-3r\dot r^2+(r+2\mh ){(r-\mh )^2 \over (r+\mh )^2}}\\
\mathcenter{\ms15 + (r-\mh)(r^2+4r\mh+\mh^2)\Omega^2\Big] }\\
\mathflushright{\ms15 - {\mc \over \rc^3} 2(r^2-\mh^2){dP \over d\theta} \dot\theta\dot r}\\
\\
\mathflushleft{\ddot\theta = {4\mh-2r \over r^2-\mh^2}\dot\theta\dot r + \cos\theta\sin\theta \dot\phi^2}\\
\mathonefourth{\ms15 + {\mc \over \rc^3} P [-2(\mh+2r)\dot\theta\dot r]}\\
\mathcenter{\ms15 + {\mc \over \rc^3} {dP \over d\theta} {1 \over 2} \Big[-(3r^2+2r\mh-3\mh^2)\dot\theta^2}\\
\mathflushright{\ms15 + (r^2+2r\mh-\mh^2)\sin^2\theta \ \dot\phi^2 + {(r-\mh)^2 \over (r+\mh)^2} + \dot r^2\Big]}\\
\\
\mathflushleft{\ddot\phi \ =\ {4\mh-2r \over r^2-\mh^2} \dot\phi\dot r - 2 \cot\theta \ \dot\theta\dot\phi - 2{\mc \over \rc^3} P(\mh+2r)\dot\phi\dot r}\\
\mathflushright{\ms15 -2{\mc \over \rc^3} {dP \over d\theta} (r^2+r\mh-\mh^2)\dot\theta\dot\phi \, ,}\\
\end{array}
\end{equation}
where dot denotes differentiation with respect to Schwarzschild coordinate time, 
and $\Omega^2 = \dot\theta^2 + \sin^2\theta\dot\phi^2$.
To avoid coordinate singularities, it is convenient to integrate (\ref{2nineteen}) in 
Cartesian coordinates.  To facilitate an otherwise tedious transformation, we first 
rewrite the equations of motion in 3-vector form.  
We make the identification 
\begin{equation}
\label{defacc}
\begin{array}{lll}
{\bf \ddot r} = & &( \ddot r-r\dot\phi^2\sin^2\theta-r\dot\theta^2){\partial\over\partial r}\\
\ms15 &+ &( \ddot\theta+(2 \dot r\dot\theta-r\dot\phi^2\sin\theta \cos\theta)/r ){\partial \over \partial\theta}\\
\ms15 &+ &( \ddot\phi +(2 \dot r\dot\phi \sin\theta+2r\dot\theta\dot\phi \cos\theta)/(r\sin\theta) ){\partial\over\partial\phi} \, ,\\
\end{array}
\end{equation}
a vector which we have constructed to be the same as the Newtonian 3-acceleration in spherical 
coordinates and we know transforms to $(\ddot x,\ddot y,\ddot z)$ in Cartesian coordinates.  Identifying   
the velocity 3-vector ${\bf v} = (\dot r,\dot\theta,\dot\phi)$ in the spherical 
coordinate basis and setting ${\bf r} = r{\partial\over\partial r}$ allows us to write  
(\ref{2nineteen}) in the compact form 3-vector form
\begin{equation}
\label{rvector}
\begin{array}{l}
  \displaystyle
\mathflushleft{{\bf \ddot r} = -{\mh \over r^2} [A{\bf e_r} + B{\bf v}]} \\
  \displaystyle
\mathflushright{\ms15 + {\mc \over \rc^3 } \left[\alpha \left({ 3{\bf r}\cdot {\bf \rc} \over \rc^2 }{\bf \rc} - {\bf r}\right) + \beta {\bf v} + \gamma {\bf e_r} \right] \, ,}\\
\end{array}
\end{equation}
where $A$ and $B$ are again given by Eqs.~(\ref{twelve}) and ~(\ref{thirteen}).  
The quantities $\alpha$, $\beta$, and $\gamma$ are given by
\begin{equation}
\label{2twentyone}
\begin{array}{l}
  \displaystyle
\mathflushleft{\alpha =  (r^2+2r\mh-\mh^2){v^2-\dot r^2 \over r^2} + \dot r^2 + {(r-\mh)^2 \over (r+\mh)^2}} \\
\\
  \displaystyle
\mathflushleft{\beta = -2(3\cos^2\Theta-1)(\mh+2r)\dot r} \\
  \displaystyle
\mathflushright{\ms15 + 12\cos\Theta\sin\Theta(r^2+r\mh-\mh^2)\dot\Theta}\\
\\
  \displaystyle
\mathflushleft{\gamma = (3\cos^2\Theta-1)\Biggl[2\mh\bigl(\dot r^2+{(r-\mh)^2 \over (r+\mh)^2} \bigr)}\\ 
  \displaystyle
\maththreefourths{\ms15 + \mh(r^2-2r\mh-\mh^2) {v^2-\dot r^2 \over r^2} \Biggr]}\\
  \displaystyle
\mathflushright{\ms15 - 12r\mh \dot r\dot\Theta\cos\Theta\sin\Theta \, ,}\\
\end{array}
\end{equation}
where $\Theta$ is the angle between $\bf r$ and ${\bf \rc}$.  It is now straightforward
to write out the Cartesian components of Eq.~(\ref{rvector}).  
To incorporate the (slow) inspiral of the
companion in the context of our quasi-static approximation, we treat ${\bf \rc} = {\bf \rc(t)}$ 
in Eq. (\ref{rvector}) as a 
parameter, which is slowly evolved in accord with 
Eqs. (\ref{fifteen}), (\ref{sixteen}) and (\ref{seventeen}).

The leading terms independent of the companion $\mc$ are identical in
both (\ref{rvector}) and (\ref{six}), a result
which is consistent with the adoption of hybrid equations 
in the PN analysis. The tidal term in  
(\ref{rvector}) reduces to the lowest-order Newtonian expression used in
(\ref{six}) in the weak-field, slow-velocity limit where 
$ v \ll 1$, $ \mh/r \ll 1$. Since the test particles orbit close to the hole, the
relativistic tidal expression will indeed cause departures from the motion 
predicted by the Newtonian tidal term. 

\subsection{Numerical Implementation}

A spherical swarm identical to the one evolved with the hybrid-PN equations
of motion was evolved with Eq.(\ref{rvector}).  Because collapse occurs later 
in this simulation, the swarm was evolved until $\rc/M=26.30$.  The results are summarized 
in Figures 3 -- 5.  
In Figure 3, the mean cluster radius is computed once again and compared to 
the hybrid-PN result.  The qualitative nature of the 
evolution is the same, and the existence of a crushing instability is evident.
The main effect of the new relativistic terms is to delay the 
collapse until the perturber gets somewhat closer to the hole.
In particular, of 20,000 particles, 16,324 fell into the hole (81.6 \%) by the time
the companion reached $\rc/\mh = 26.30$, a slightly smaller percentage than in the hybrid-PN simulation.  
Once again, the compact cluster is observed to be stable in isolation, but 
driven to collapse by the presence of a sufficiently strong tidal perturbation.

\begin{figure}
\epsfxsize=3in
\begin{center}
\leavevmode
\epsffile{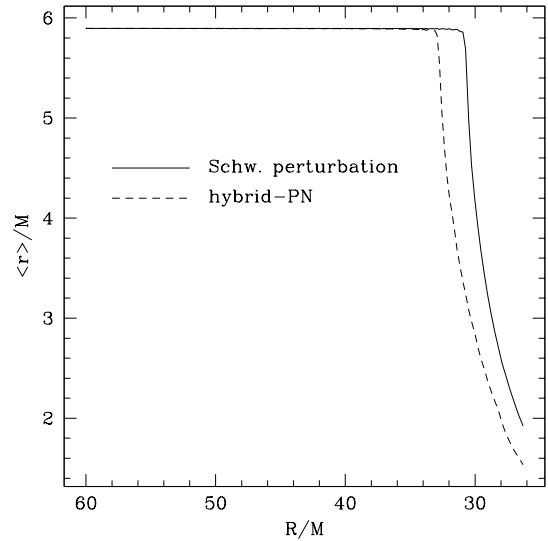}
\end{center}
\epsfxsize=3in
\begin{center}
\leavevmode
\epsffile{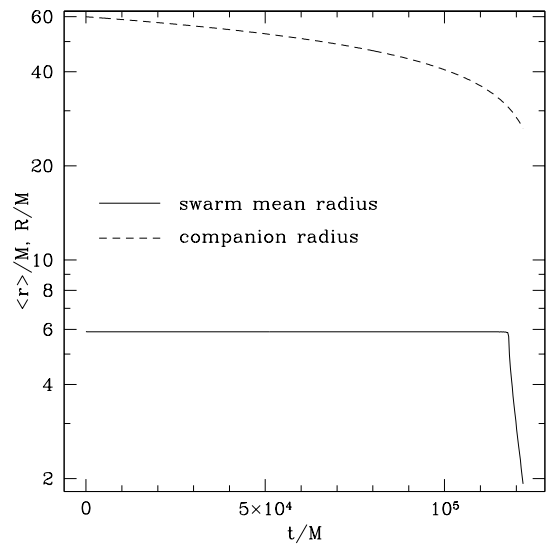}
\end{center}
\caption{Evolution of the cluster in the Schwarzschild perturbation treatment.
The mean radius of the 20,000 test-particles in the initially spherical swarm
is plotted as a function of the tidal companion radius in the top figure and
as a function of coordinate time in the bottom figure. In the top figure we
compare the results found for our two treatments; the behavior is qualitatively
similar, but the crushing effect occurs slightly later in the fully relativistic,
Schwarzschild perturbation treatment.}
\end{figure}

\onecolumn
\begin{figure}[p]
\begin{center}
\leavevmode
\hbox{\vbox{
	\hbox{\hbox to 3in{\hfil \Huge \bf A \hfil} \hbox to 1in{} \hbox to 3in{\hfil \Huge \bf B \hfil}}
	\hbox{\vrule width 0pt height 5pt}
	\hbox{\fbox{\epsfxsize=3in\epsffile{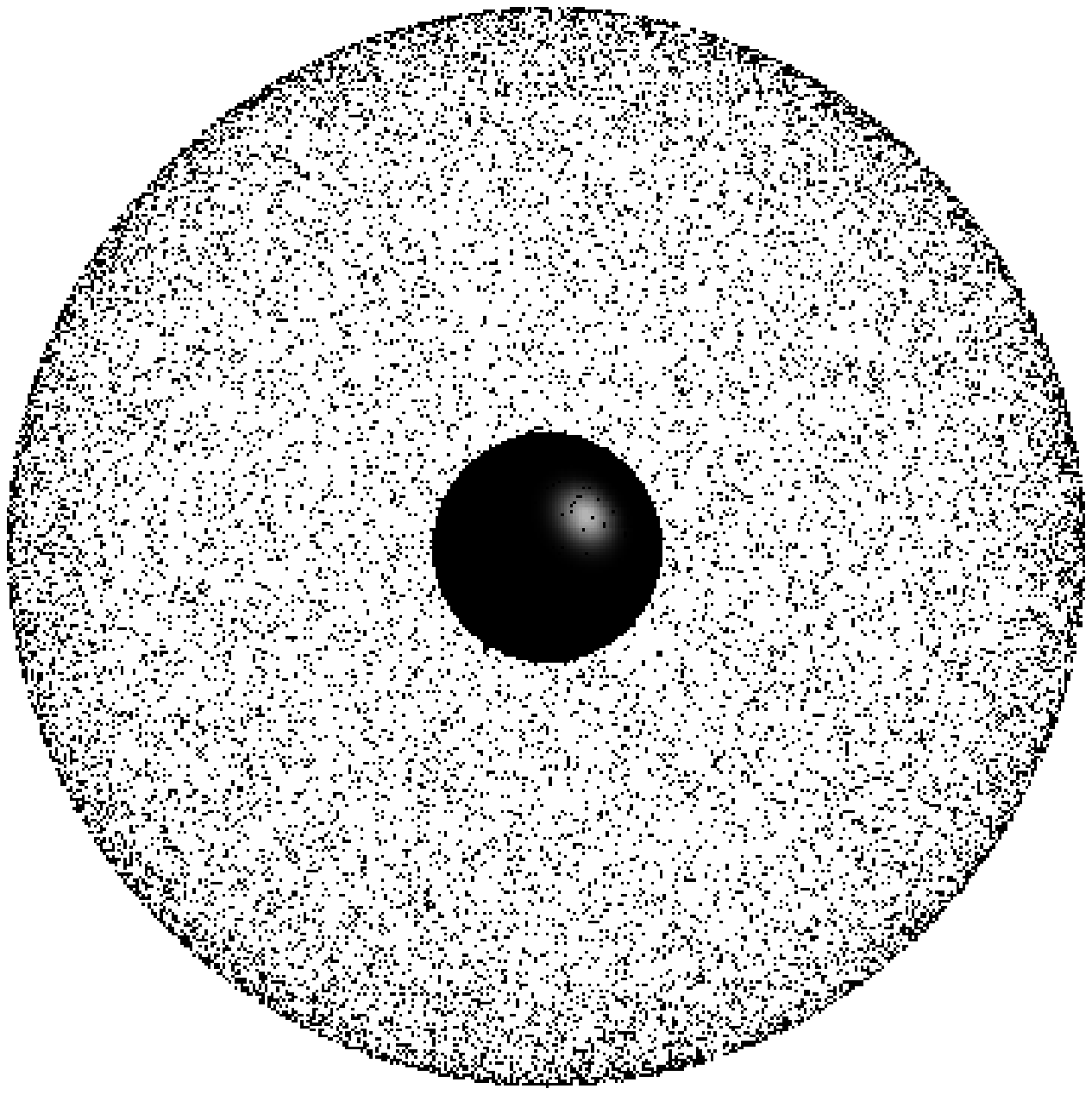}
		    \hbox to 1in{}
		    \epsfxsize=3in\epsffile{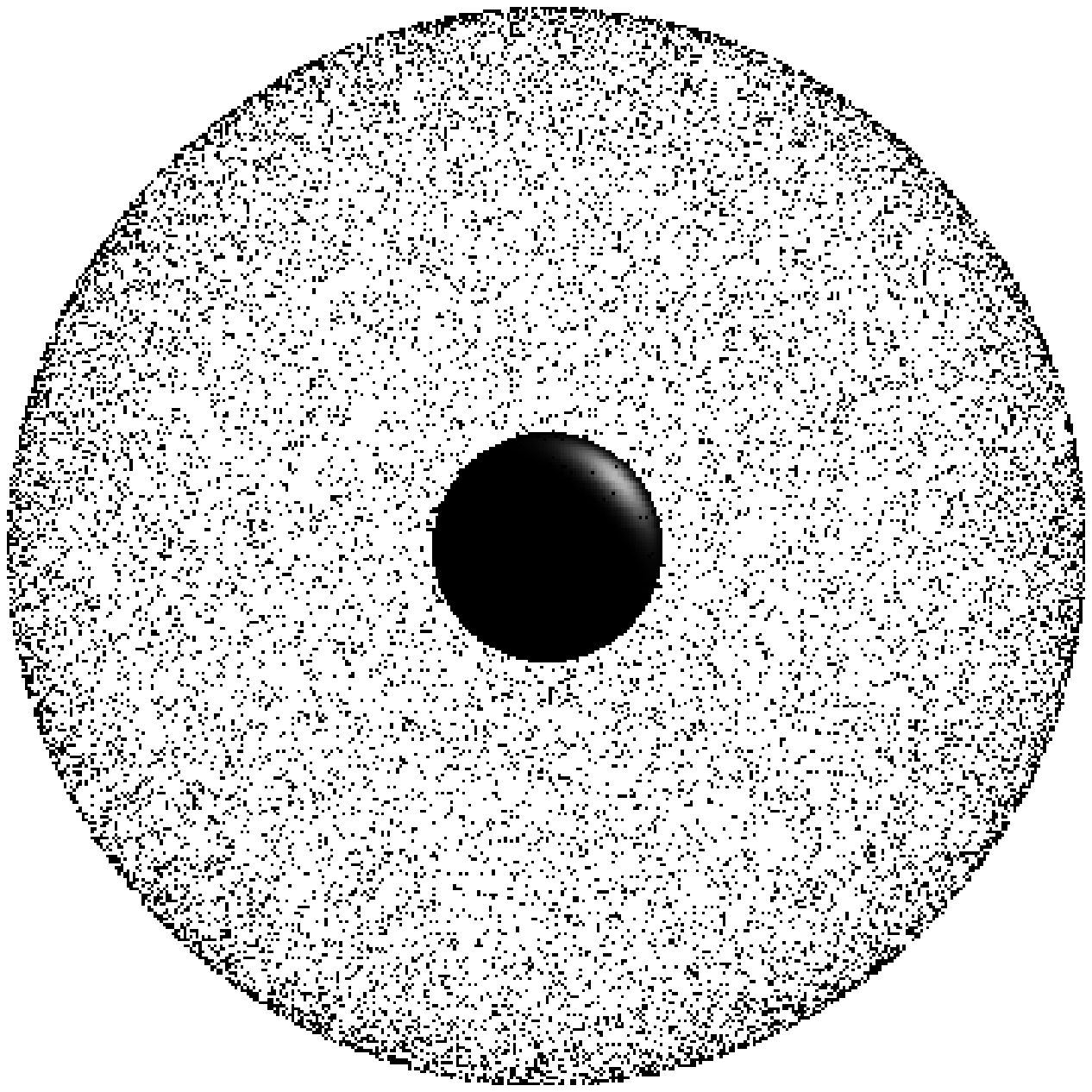}}
	      \hskip-7in
	      \vbox to 2.25in{\vskip5pt \hbox to 7in{\hfil \LARGE $t/\mh = 0$ \hfil} \vfil}
	}
	\hbox{\fbox{\epsfxsize=3in\epsffile{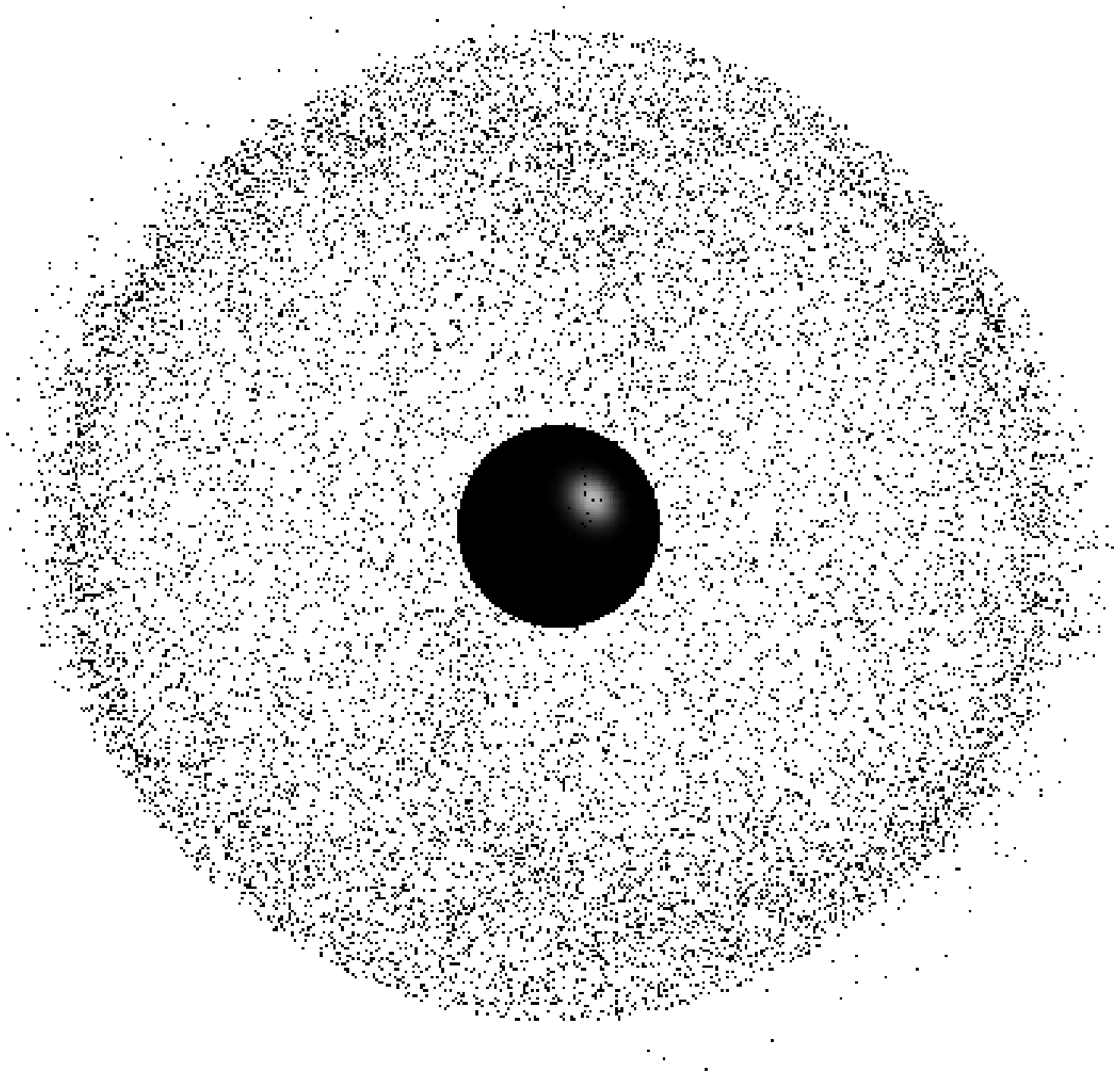}
		    \hbox to 1in{}
		    \epsfxsize=3in\epsffile{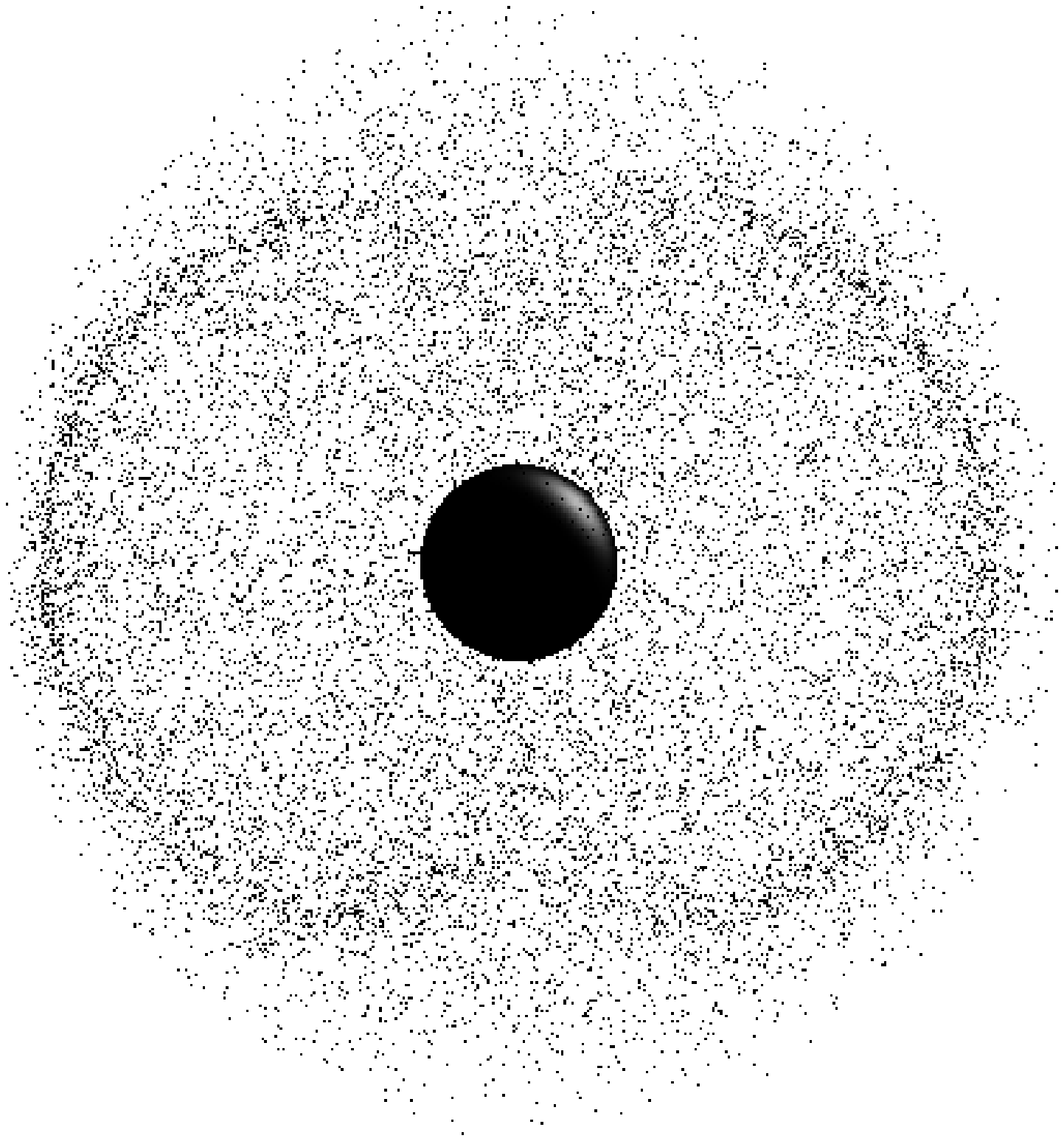}}
	      \hskip-7in
	      \vbox to 2.25in{\vskip5pt \hbox to 7in{\hfil \LARGE $t/\mh = 1.18\times10^5$ \hfil} \vfil}
	}
	\hbox{\fbox{\epsfxsize=3in\epsffile{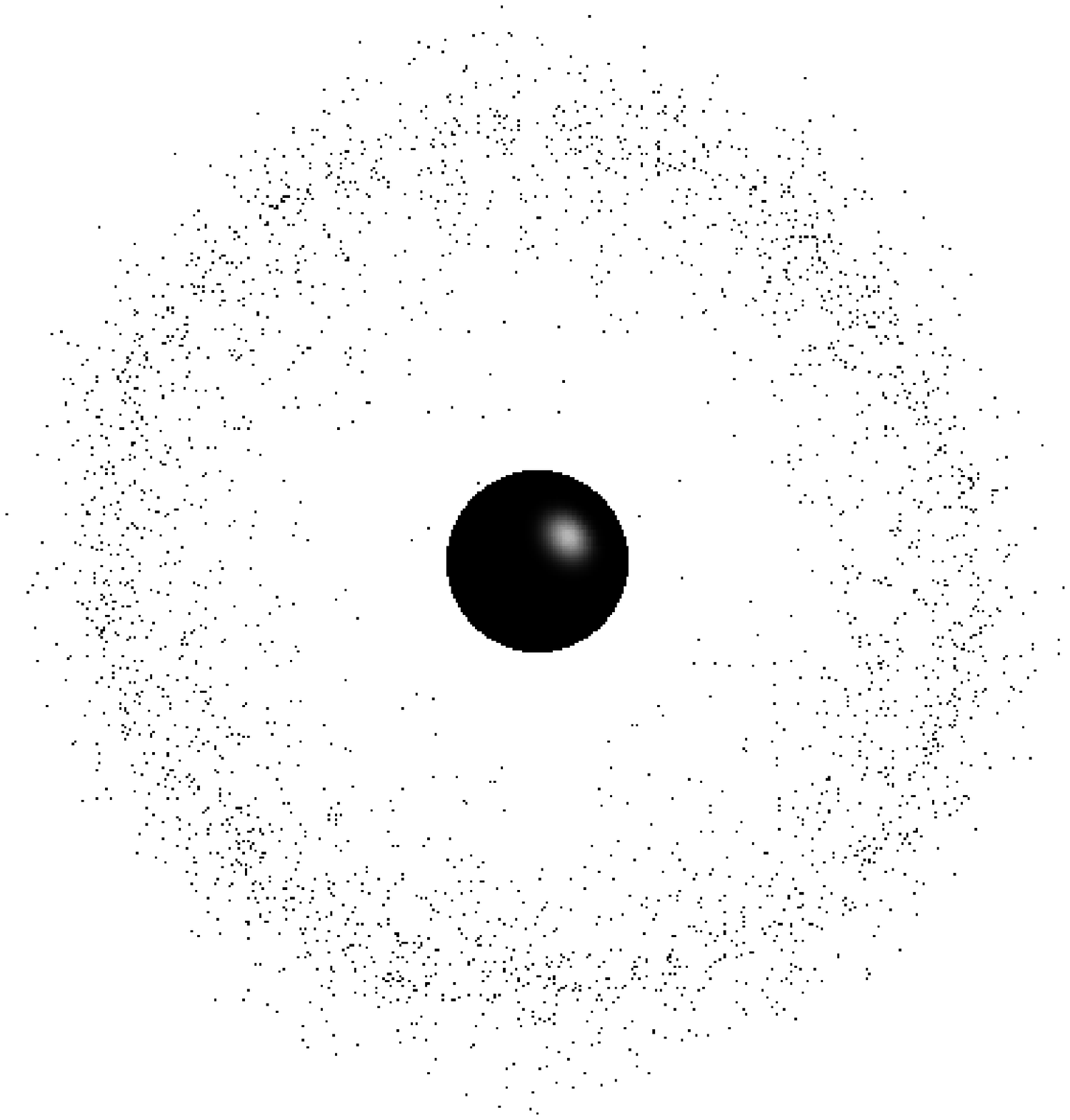}
		    \hbox to 1in{}
		    \epsfxsize=3in\epsffile{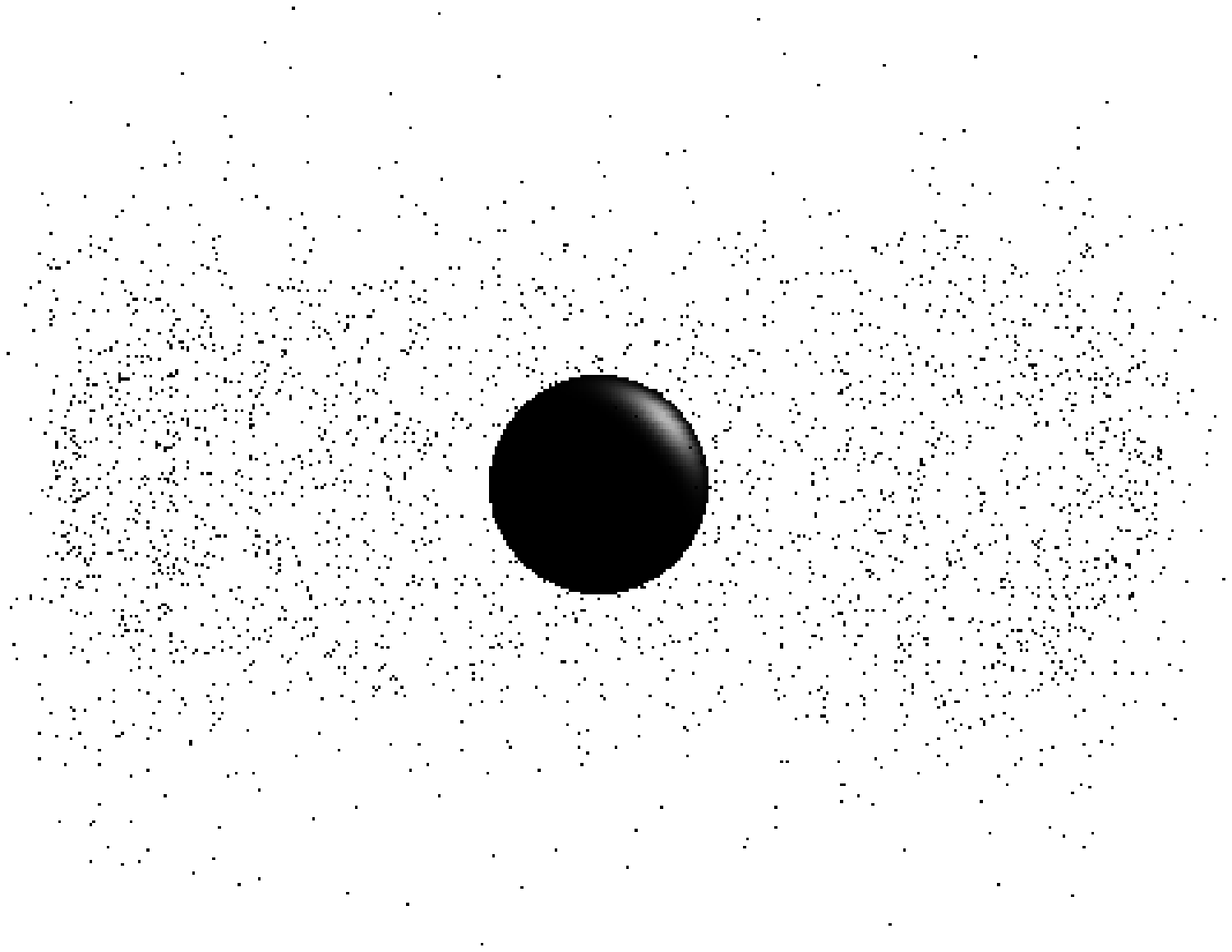}}
	      \hskip-7in
	      \vbox to 2.25in{\vskip5pt \hbox to 7in{\hfil \LARGE $t/\mh = 1.22\times10^5$ \hfil} \vfil}
	}
	\hbox{\vrule width 0pt height 5pt}
}}
\end{center}
\caption{Snapshots of the cluster swarm at selected times during the binary inspiral.
In $A$ we show the view looking down along the $z$-axis, which is perpendicular to
the orbital plane of the companion; in $B$ we show the view looking along the $x$-axis, 
which lies in the  orbital plane of the companion. For the
top frame the companion is at $\rc/\mh = 60$; for the middle frame the companion
is at $\rc/\mh = 30.7$; for the bottom frame the companion is at $\rc/\mh = 26.30$.}
\end{figure}
\twocolumn

\begin{figure}
\begin{center}
\leavevmode
\hbox to 246pt{\hskip40pt
\hbox{\vbox{
	\hbox{\fbox{\epsfxsize=2in\epsffile{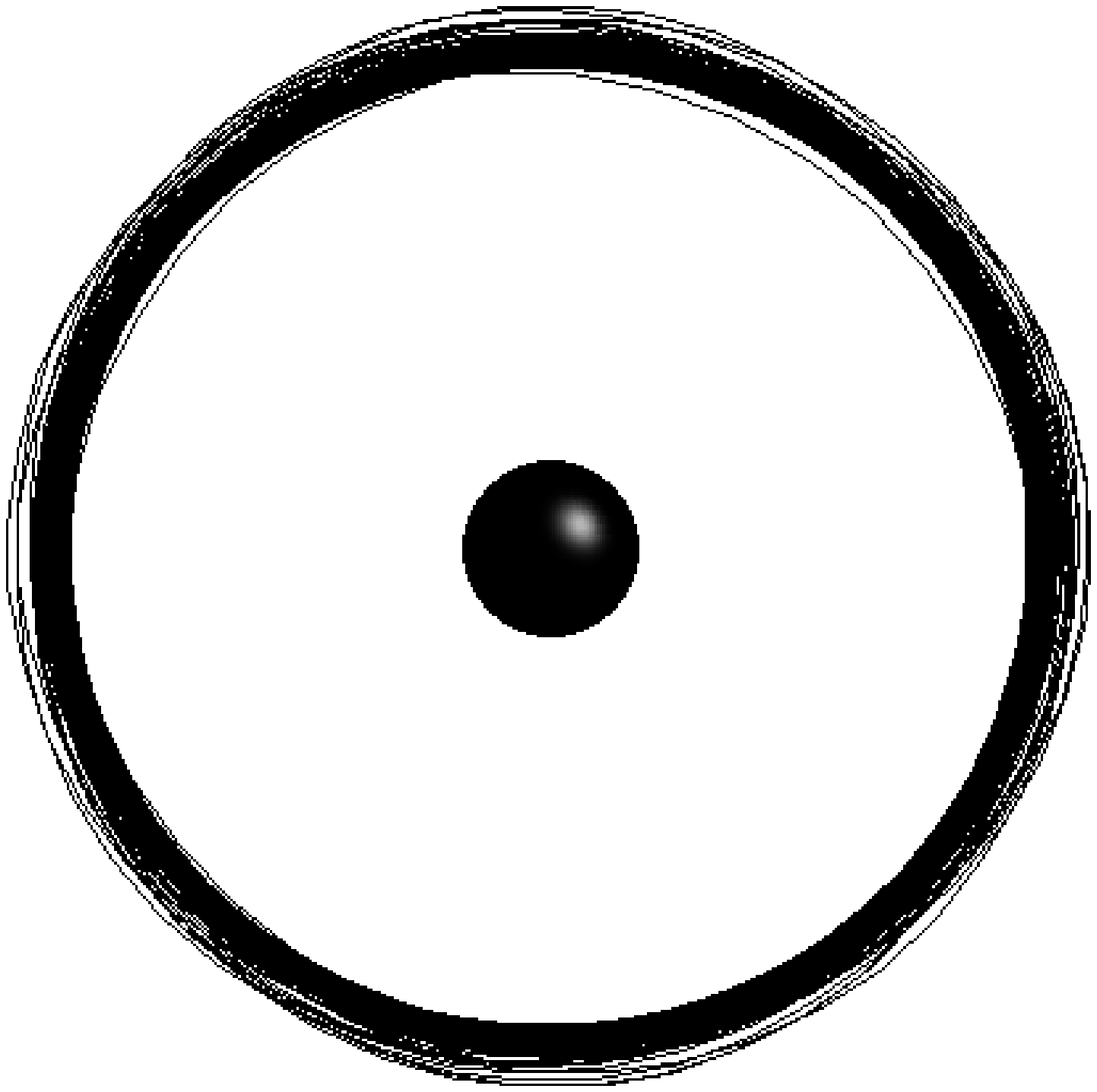}}
	      \hskip-2in
		\vbox to 1.5in{\vskip5pt \hbox{\LARGE a} \vfil}}
	\hbox{\fbox{\epsfxsize=2in\epsffile{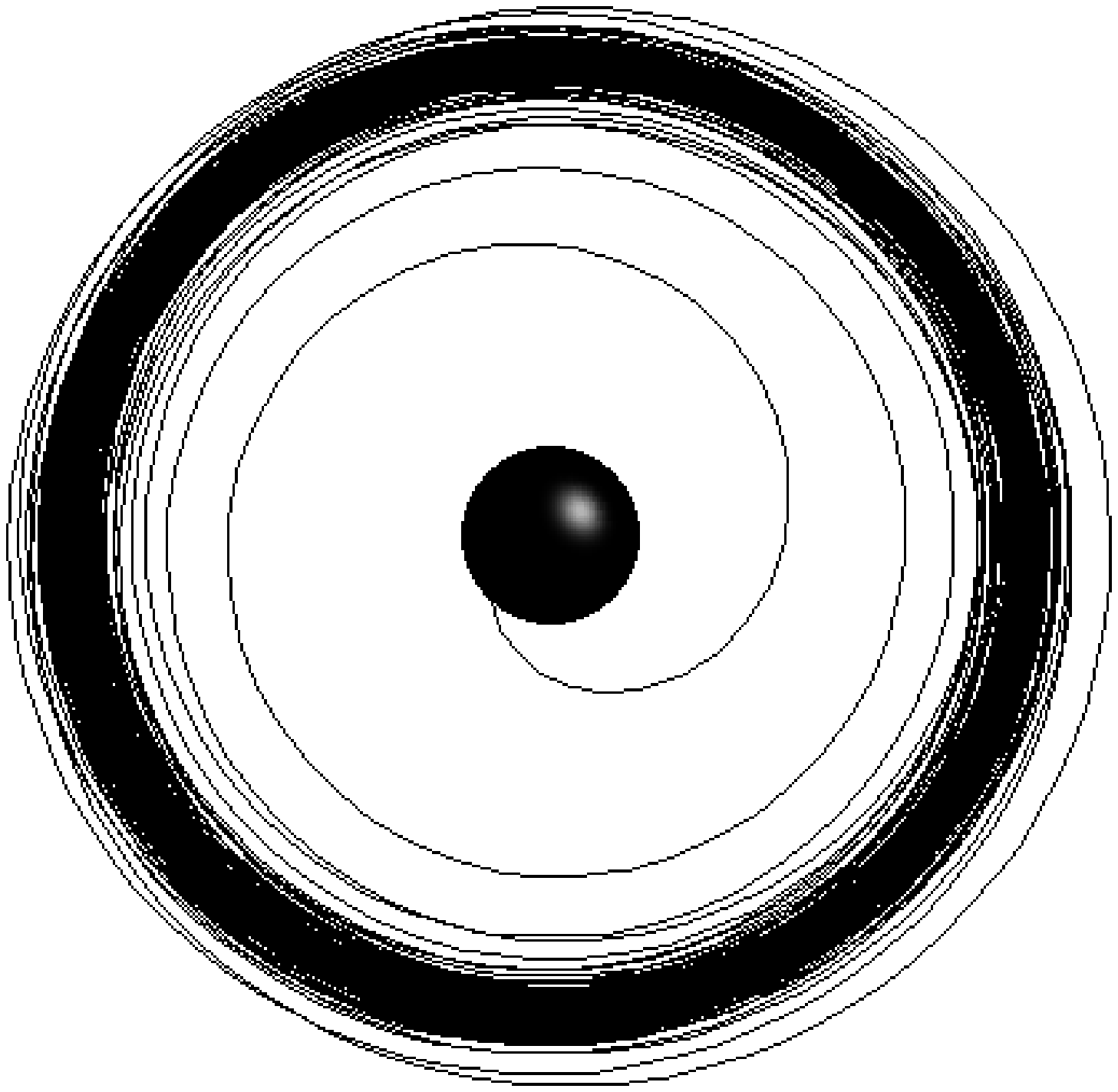}}
	      \hskip-2in
		\vbox to 1.5in{\vskip5pt \hbox{\LARGE b} \vfil}}
	\hbox{\fbox{\epsfxsize=2in\epsffile{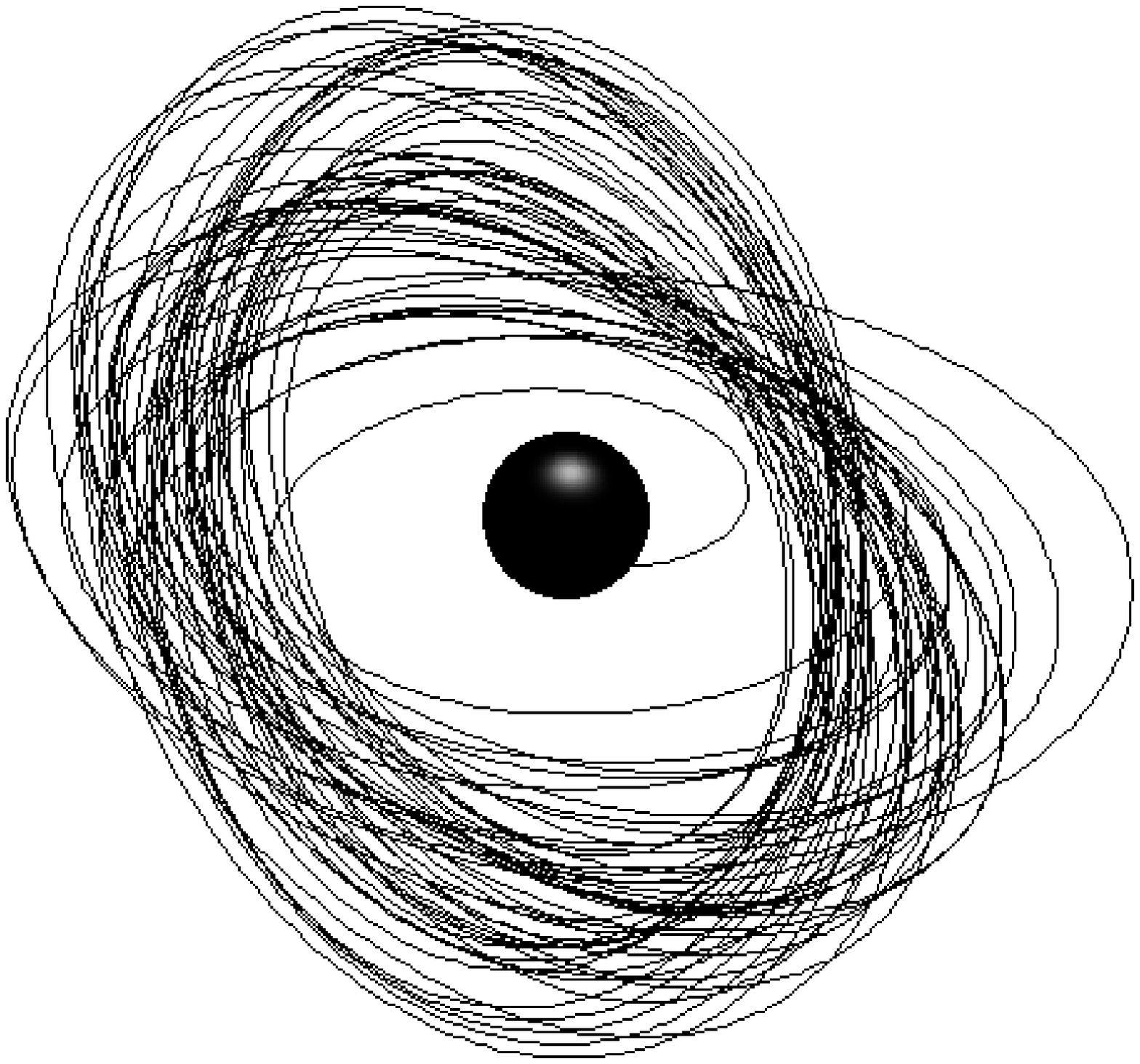}}
	      \hskip-2in
		\vbox to 1.5in{\vskip5pt \hbox{\LARGE c} \vfil}}
	\hbox{\vrule width 0pt height 5pt}
}}
\hfil}
\end{center}
\caption{ Test-particle trajectories about the black hole during the
inspiral of the binary companion from $\rc/\mh=60$ to $\rc/\mh=26.30$.
Frame (a) shows the orbit of a particle moving in the orbital plane
of the companion; it is not captured by the time the integrations terminate.
Frame (b) shows the capture of a particle initially orbiting in
a plane perpendicular to the companion plane. Frame (c)
shows the precession and capture of a particle initially orbiting at
an angle of $45^o$ to the companion plane.}
\end{figure}

As the geodesic equations in the perturbative relativistic treatment are
derived from a self-consistent Lagrangian, they satisfy strict conservation laws
even in the presence of a companion, assuming it is stationary. These conservation
laws provide a means of testing our code and particle integration scheme.
For example, if the companion is fixed at an arbitrary position on the 
z-axis, the perturbed Schwarzschild spacetime (\ref{2fifteen}) admits two Killing vectors,
$\partial\over\partial t $ and $\partial\over\partial \phi$, yielding 
conservation of particle energy $p_t$ and angular momentum $p_{\phi}$,
even for large tidal fields. 
Given that we linearize the tidal field and retain only the lowest order terms in $\mc$ in our
equations of motion (\ref{rvector}), 
energy and angular momentum conservation are no longer exact. However, we have
tested our code and have shown that it reliably obeys 
these conservation laws to the required order.
 
The evolution of the swarm is depicted in Figure 4, where snapshots of the
cluster are  shown at three different times from two different viewing angles.
By the end of the simulation, 
the original spherical swarm is reduced to a sparse cylindrical band of particles whose
axis is perpendicular to the orbital plane of the companion.
Apparently, particle orbits at small inclination angles to this plane are more stable 
than those which are perpendicular, a result already noted in
~\cite{Us}.  This feature is evident in Figure 5, which plots the trajectories of three 
representative particles.  The simulation confirms that gravitational collapse can be induced
in a collisionless cluster by the tidal field of a binary companion.

\appendix
\section*{Adiabatic Invariance of the Horizon Area}
Bekenstein~\cite{Bek} has argued recently 
that the horizon area of a near-equilibrium black hole is an 
adiabatic invariant.  Using the focusing equation, he derives the following conditions which must 
be satisfied in order for area to remain fixed:
\begin{equation}
\label{condit}
 T_{\alpha\beta}l^{\alpha}l^{\beta} = 0; \ \ \ \
 C_{\alpha\beta\gamma\delta}l^{\alpha}m^{\beta}l^{\gamma}m^{\delta} = 0 \, , \ \
 \mbox{(on the horizon)}
\end{equation}
where $C_{\alpha\beta\gamma\delta}$ is the Weyl conformal tensor, and $l^{\alpha}$ and $m^{\alpha}$ 
are legs of the Newman-Penrose tetrad ($l^{\alpha}$ are tangents to the null generators of the 
event horizon).  For our quasi-static scenario in which the companion undergoes a very slow
inward spiral,
there is no matter and and essentially no wave flux at the horizon, so both 
conditions (\ref{condit}) hold.  (Other situations in which adiabatic 
invariance has been demonstrated are discussed in~\cite{Bek} and~\cite{Mayo}.)  We may use our 
metric in Schwarzschild coordinates (\ref{2fourteen}) to verify explicitly 
that in accord with Bekenstein's conjecture, the
horizon area remains invariant in 
our model as the companion is brought in slowly from large distance. 

First, rewrite (\ref{2fourteen}) in Kruskal-Szekeres coordinates, ignoring angular terms,
reintroducing areal coordinate $r$ and  
defining $\kappa = {\mc \over \rc^3}$, 
\begin{equation}
\label{krus}
\begin{array}{l}
  \displaystyle
\mathflushleft{ds^2 = {32\mh^3 \over r}e^{-r/2\mh}\Bigg\{-dv^2 + du^2}\\
  \displaystyle
\mathflushright{\ms15 + {2P\kappa r \over \mh}e^{-r/2\mh}\Big[(u^2+v^2)(du^2+dv^2)-4uv\,du\,dv\Big]\Bigg\} \, .}\\
\end{array}
\end{equation}
Note that this metric agrees with that derived by Vishveshwara~\cite{Vis} for general even 
perturbations of Schwarzschild in Kruskal coordinates.  Null geodesics are found by setting 
$ds^2 = 0$, for which
\begin{equation}
\label{null}
\begin{array}{l}
\displaystyle
\mathflushleft{{du \over dv} = { {8uv\kappa Pr \over \mh}e^{-r/2\mh} \over 2(1+(u^2+v^2){2\kappa Pr \over \mh}e^{-r/2\mh}) }}\\
\displaystyle
\mathflushright{\mathstrutter{35pt} \pm { \sqrt{ {(8uv\kappa Pr)^2 \over \mh^2}e^{-r/\mh}-4\Big[(u^2+v^2)^2{(2\kappa Pr)^2 \over \mh^2}
e^{-r/\mh}-1\Big]} \over 2\Big[1+(u^2+v^2){2\kappa Pr \over \mh}e^{-r/2\mh}\Big] } \, .} \\
\end{array}
\end{equation}
Direct substitution verifies that (\ref{null}) is satisfied by $u = \pm v$, $r(t) = 2\mh$.  Therefore, 
the horizon on the perturbed black hole remains at $2\mh$.
The surface area of a shell of constant radius is given by
\begin{equation}
\label{integral}
 A = \int_0^{2\pi}\!\!\!\!\!\int_0^{\pi} \sqrt{g_{\theta\theta}g_{\phi\phi}} d\theta d\phi = 4\pi r^2 \, .
\end{equation}
Due to the orthogonality of  Legendre polynomials ($P_2$ and $P_0$ in this case), the 
quadrupole perturbations in the angular piece of metric (\ref{2fourteen}) do not change 
areas of shells.  Therefore, the area of 
the spherical shell $r=2\mh$, the location of the horizon, remains constant.

Hartle ~\cite{Hart} has 
studied the effect of tidal fields from more general stationary sources on the 
horizon area of 
slowly rotating black holes.  In the limit of a nonrotating black hole, his 
analysis gives a vanishing time rate of change for the area, consistent with our
explicit calculation above.

\vskip20pt
\centerline{\bf ACKNOWLEDGMENTS}
\vskip15pt
It is a pleasure to thank A.M. Abrahams, T.W. Baumgarte and N.J. Cornish
for stimulating discussions and useful suggestions.  We are also grateful to 
K.~Alvi, L.~Burko, and Y.T.~Liu for pointing out to us a typo in equation
(\ref{sixteen}) in an earlier draft.
The Undergraduate Research Team in theoretical astrophysics and general relativity
(M. Duez, E. Engelhard, J. Fregeau and K. Huffenberger) gratefully
acknowledges support from the Departments of Physics and Astronomy at the University of Illinois
at Urbana-Champaign (UIUC). Much of the calculation and visualization were performed at 
the National Center for Supercomputing Applications at UIUC.
This paper was supported in part by NSF Grant AST 96-18524 and NASA
Grant NAG 5-7152 to UIUC.


\begin{references}
\bibitem{OUR} Our inspiral approximation is performed in the
same  spirit as the hybrid approximation, where complete PN
consistency is sacrificed in order to insure  that the
equations recover some essential dynamical behavior  arising
at higher order.
\end{references}
\end{document}